# Plasmonic waveguides cladded by hyperbolic metamaterials

Satoshi Ishii,[1,2] Mikhail Y. Shalaginov,[1] Viktoriia E. Babicheva,[1,3,4] Alexandra Boltasseva,[1,3] and Alexander V. Kildishev[1]*

[1] Birck Nanotechnology Center and School of Electrical and Computer Engineering, Purdue University, West Lafayette, IN 47907, USA
[2] International Center for Materials Nanoarchitectonics (MANA), National Institute for Materials Science (NIMS), Tsukuba, Ibaraki 305-0044, Japan
[3] DTU Fotonik – Department of Photonics Engineering, Technical University of Denmark, Oersteds Plads 343, Kgs. Lyngby 2800, Denmark
[4] ITMO University, Kronverkskiy, 49, St. Petersburg 197101, Russia
*kildishev@purdue.edu

Strongly anisotropic media with hyperbolic dispersion can be used for claddings of plasmonic waveguides. In order to analyze the fundamental properties of such waveguides, we analytically study 1D waveguides arranged of a hyperbolic metamaterial (HMM) in a HMM-Insulator-HMM (HIH) structure. We show that hyperbolic metamaterial claddings give flexibility in designing the properties of HIH waveguides. Our comparative study on 1D plasmonic waveguides reveals that HIH-type waveguides can have a higher performance than MIM or IMI waveguides.

Plasmonic waveguides (PWs), guiding surface plasmon polaritons (SPPs) at the nanoscale, are the basis of optical nano-circuits. For example, the finite-length PW segments with subwavelength apertures are used as phase-controlling elements in metallic diffraction lenses [1-5].

Although 2D PWs have tighter confinement and are more practical, here we limit ourselves to a theoretical study of 1D PWs in order to present the properties of guided SPPs more clearly. Basic 1D PW are MIM and IMI structures where M and I stand for metal and insulator, respectively. MIM and IMI PWs have complementary guiding properties, i.e. the MIM design has better confinement but shorter propagation length than IMI [6].

To enhance the functionality of standard 1D PWs, a promising route could be the use of anisotropic materials. Among them, strongly anisotropic materials with hyperbolic dispersions are of particular interests [7]. The unique properties of hyperbolic metamaterials (HMMs) have drawn significant attention and have been utilized for various applications, such as subwavelength imaging [8-11], light compression [12] and subwavelength interferences [13]. The HMMs have already been studied as the cores and claddings of photonic waveguides. Some of the past studies of the photonic modes in anisotropic waveguides include waveguides with a large mode index [14, 15], negative refractive index [16, 17], and slow light [18, 19].

As opposed to the previous studies, in the present work we consider 1D PWs in which the claddings are HMMs where translational symmetries are broken [20, 21]. Our guideline would help to choose the best PWs in terms of functionality. We analytically study fundamental guiding properties of HIH PWs using realistic materials and compare HIH PWs against MIM and IMI PWs. Our analysis aims at a basic understanding of HIH PWs and could be extended to designing 2D PWs cladded by HMMs.

Metal-dielectric lamellar structures are the simplest HMMs [22], where the effective permittivities of perpendicular ($\varepsilon_\perp$) and parallel components ($\varepsilon_\parallel$) can be estimated through the effective media theory (EMT), $\varepsilon_\perp^{-1} = (1-r)\varepsilon_d^{-1} + r\varepsilon_m^{-1}$, $\varepsilon_\parallel = r\varepsilon_m + (1-r)\varepsilon_d$, where $\varepsilon_m$ and $\varepsilon_d$ are the permittivities of metal and dielectric, respectively, and $r$ is the ratio of the metal layer thickness to the period of the lamellar HMM. All the calculations shown in the paper are done by taking the effective permittivities of the lamellar structures. However, we note that the differences between the exact solutions are negligible within the conditions analyzed in this paper (results not presented).

To begin, we derive a dispersion equation (DE) of 1D PWs that have different uniaxial anisotropy for the top cladding, the core, and the bottom cladding. The derivation and the final DE are shown in the Appendix. The DE becomes simpler if both claddings are identical. The DEs for plasmonic guided modes can be easily modified for photonic guided modes (see Appendix).

Here, we analytically calculate the properties of HIH PWs having isotropic dielectric cores ($\bar{\varepsilon} = \text{diag}(\varepsilon_c, \varepsilon_c, \varepsilon_c)$) using the DE. When constructing HIH PWs, the optical axes of HMMs are always perpendicular to the propagation direction so that the parallel component ($\varepsilon_\parallel$) is negative. This condition ensures plasmonic propagation. Note that further on we consider only long-range propagating modes defined by Eq. A(9), as we are not interested in short-range modes given by Eq. A(10).

To construct lamellar HMMs we take gold ($\varepsilon_m = \varepsilon_{Au}$) [23] and silica ($\varepsilon_d = 2.25$) for metal and dielectric, respectively, and consider different metal ratios ($r$ = 0.2, 0.5, 0.8 and 1). As shown in Fig. 1, the effective permittivities provide hyperbolic regime for all $r$ (except for $r$ = 1, when the cladding consists only of gold and the waveguide becomes equivalent to an MIM PW).

Using the HMMs with the permittivities plotted in Fig. 1, we then plot the guiding properties of the HMM PWs in Fig. 2: magnetic field amplitude, mode index $n_{\text{eff}} = \text{Re}[k_x](k_0)^{-1}$, propagation length, $L = \text{Im}[2k_x]^{-1}$, and penetration depth on the one side of the cladding $\delta = \text{Im}[2k_y]^{-1}$. The permittivity of the core ($\varepsilon_c$) and the core thickness ($2d$) are fixed to unity and 50 nm, respectively. The magnetic field plot show plasmonic propagations, which are similar to MIM PWs. The effective indexes are around 1.45 and depend slightly on the changes in $r$. In near-IR region, the propagation lengths and the penetration depths are around 8 μm and $20 - 50$ nm, respectively, except $r = 0.2$. At $r = 0.2$, the propagation lengths and the penetration depths are around 10 μm and 100 nm, respectively.

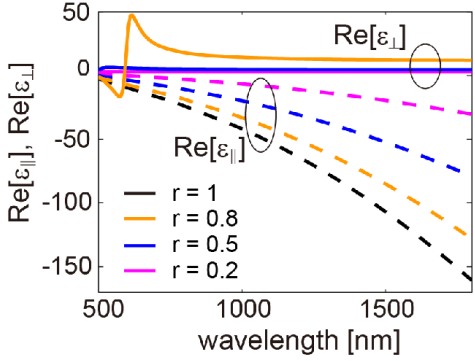

Fig. 1. Effective anisotropic permittivities of the lamellar structures for $r = 0.2, 0.5, 0.8$ and $1$. The solid lines and dashed lines represent perpendicular and parallel components, respectively.

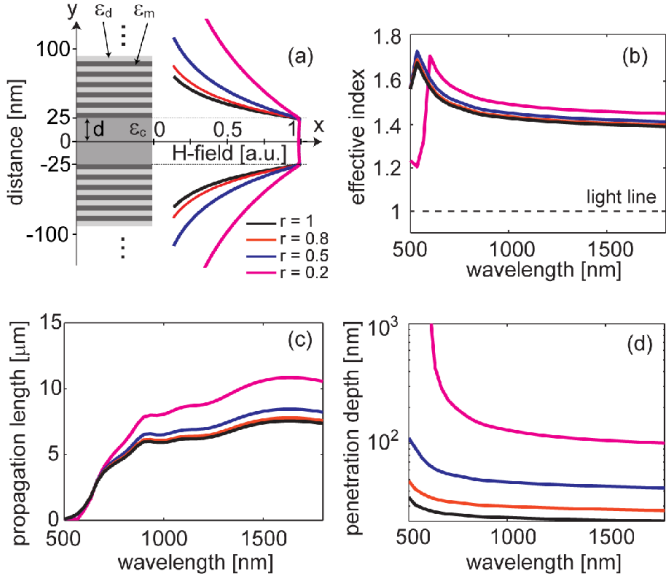

Fig. 2. (a) Schematic of a symmetric HIH PW. Magnetic field amplitudes along the y-axis at 1550 nm are superimposed. (b) Effective index, (c) propagation length and (d) penetration depth of the HIH PWs at $r = 0.2, 0.5, 0.8$ and $1$.

Next, we plot the core-thickness dependences of four different PWs: MIM, IMI and two different HIH PWs in Fig. 3. It is well known that propagation length becomes shorter as the core thickness becomes smaller for MIM PWs, while IMI PWs have the opposite dependence (see Figs. 3(a) and (b)). We need to pay attention to the HIH PWs in Figs. 3(c) and (d), while the HIH PWs at $\varepsilon_m = \varepsilon_{Au}$, $\varepsilon_d = 2.25$, $\varepsilon_c = 1$ and $r = 0.5$ maintains MIM property, the HIH PWs at $\varepsilon_m = \varepsilon_{Au}$, $\varepsilon_d = 2.25$, $\varepsilon_c = 1.88$ (permittivity of MgF$_2$) and $r = 0.2$ have a property of IMI PWs.

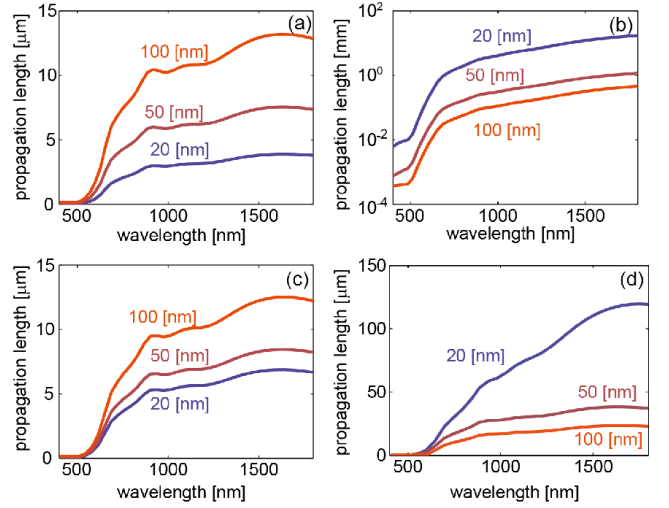

Fig. 3. Propagation length of (a) MIM, (b) IMI and (c, d) HIH PWs for three difference core thicknessess (20, 50 and 100 nm). For HIH PWs, (c) $\varepsilon_m = \varepsilon_{Au}$, $\varepsilon_d = 2.25$, $\varepsilon_s = 1$ and $r = 0.5$ and (d) $\varepsilon_m = \varepsilon_{Au}$, $\varepsilon_d = 2.25$, $\varepsilon_c = 1.88$ and $r = 0.2$.

After studying the properties of HIH PWs, we then compare their characteristics with two other standard PW layouts: MIM and IMI. For this purpose we introduce a figure of merit (FoM) proposed by Berini [24], $FoM = L/D$, where $D = 2(\delta + d)$ is the mode size with $\delta$ and $d$ representing the penetration depth and one-half of the core thickness, respectively, and $L$ is the propagation length. In general, it is desirable for a waveguide to have a longer propagation length and a smaller penetration depth, which results in a large FoM. The definition of FoM includes the core thickness in order to compare waveguides having different core thicknesses, however, inclusion of core thickness does not change the trends presented below.

In Fig. 4, we plot the FoMs of the HIH PWs in addition to the MIM as well as IMI PWs. The core thicknesses are fixed to 50 nm. The HMMs which are used as the claddings of HIH PWs are identical to those plotted in Fig. 1. For the particular case considered in Fig. 4 (b), the HIH PW at $r = 0.2$ have higher FoM than the MIM and IMI PWs above 1450 nm. The results indicate that making the correct choices of geometry and materials, HIH PWs can be designed to outperform MIM and IMI PWs.

The conclusion about the FoM changes with any of the following parameters: wavelength, permittivity of the constituent materials, metal ratio, and thickness of the core. To analyze the structure here we decided to limit our study by the case of telecom wavelength 1.55 μm and two materials, i.e. gold and silica. Thus, we varied $r$ and $d$ and plotted propagation length and effective mode size for the different structure as shown in Fig. 5. The vertical dotted lines show mode size which approximately corresponds to the size of conventional photonic waveguide, so this region is of particular interest. Thus, for low filling ratio $r < 0.3$, HIH PWs can have properties which are not achievable with MIM or IMI PWs. In particular, HIH waveguides can have a higher propagation length ($> 650$ μm) while maintaining a smaller mode size ($< 2$ μm).

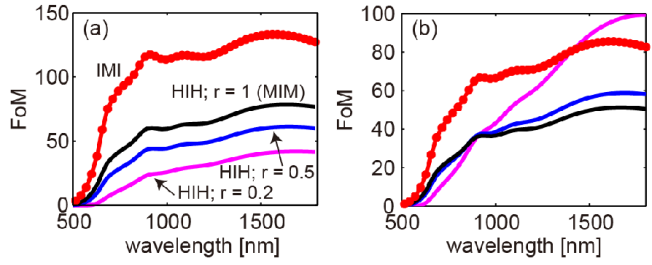

Fig. 4. FoMs of HIH PWs for various $r$ ($r$ = 0.2, 0.5, and 1) in comparison to FoM of the IMI PWs. The HIH PW at $r$ = 1 is equivalent to the MIM PW. For all cases, the core thickness is 50 nm. (a) The permittivity of the core in the HIH PWs and of the host in the IMI PW is equal to 1. (b) The permittivity of the core the HIH PWs and of the host in the IMI PW is 2.25.

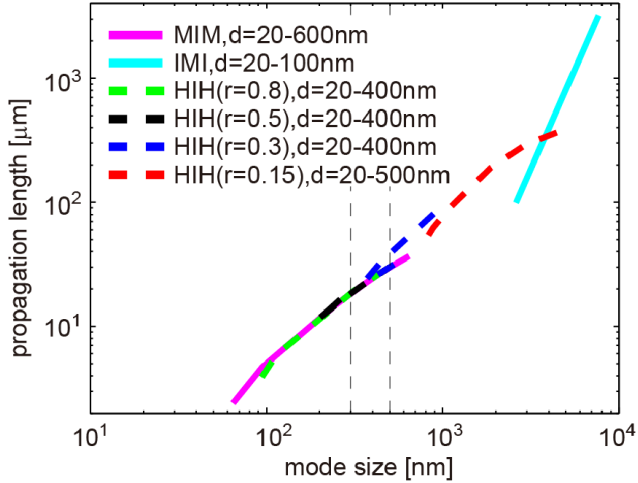

Fig. 5. Characteristics of the HIH PWs and their comparison with MIM and IMI PW layouts at the wavelength of 1.55 µm.

To summarize, We study the guiding properties of 1D PWs where the isotropic dielectric cores are cladded by HMMs. Working with HMMs enables us to tune the properties of PWs with existing materials. With a proper choice of geometry and materials, HIH PWs can have a higher performance than MIM and IMI PWs. Our current study on 1D HIH PW could be extended to 2D HIH PWs.

In practice, a critical issue for PWs is that actual metals have losses, which substantially shorten the propagation lengths. We note that using HMMs could abate the effect of losses. Moreover, passive dielectric layers of HMMs could be replaces with active materials such as dye-doped polymers [25-28], hence compensating for the propagation losses and can be applied for interconnects in on-chip devices. Further analysis will be presented elsewhere.

This work is partially supported by ARO MURI grant 56154-PH-MUR (W911NF-09-1-0539), ARO grant 57981-PH (W911NF-11-1-0359) and MRSEC NSF grant DMR-1120923 and the Strategic Information and Communications R&D Promotion Programme (SCOPE). V.E.B. acknowledges financial support from SPIE Optics and Photonics Education Scholarship, as well as Otto Mønsteds and Kaj og Hermilla Ostenfeld foundations.

**Appendix:** A schematic drawing of an 1-D anisotropic waveguide is shown in Fig. A1. The waveguide is made of *three different anisotropic layers*: an upper cladding ($A_1$), a core ($A_2$), and a lower cladding ($A_3$). The thickness of the core is $2d$; both claddings are semi-infinite in $y$-direction, range $y \in (-\infty, -d]$ defines the lower layer, while $y \in [d, \infty)$ defines the upper layer. The dielectric constants of the anisotropic materials are given as $\varepsilon_{xi}$, $\varepsilon_{yi}$, and $\varepsilon_{zi}$, where i represents the layer number, i = {1, 2, 3}. The anisotropy axes are aligned with the propagation direction and with a unit normal vector to the interface, so that $\bar{\varepsilon}_i = \mathrm{diag}(\varepsilon_{xi}, \varepsilon_{yi}, \varepsilon_{zi})$. The purpose of the study is to derive the DEs for 1-D waveguides made of anisotropic materials.

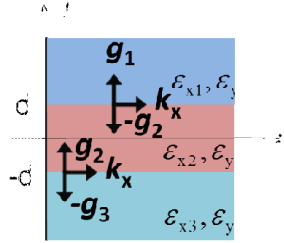

Fig. A1. Schematic drawing of an anisotropic waveguide.

To study plasmonic modes propagating along the x-axis we start from Maxwell's equations, where the time dependence factor $\exp(-i\omega t)$ is omitted for simplicity. The charge density inside the structure equals zero, so that $\nabla \cdot \mathbf{D} = \nabla \cdot \mathbf{B} = 0$. The other two equations are the following:

$$\nabla \times \mathbf{E} = -\mu_0 \partial_t (\mu \mathbf{H}), \quad \mathrm{A}(1)$$

$$\nabla \times \mathbf{H} = \varepsilon_0 \partial_t (\varepsilon \mathbf{E}), \quad \mathrm{A}(2)$$

where $\mu = 1$, and $\varepsilon = \mathrm{diag}(\varepsilon_x, \varepsilon_y, \varepsilon_z)$. From Eqs. A(1) and A(2), we can write the wave equation ($k_0 = \omega/c$),

$$\hat{\mathbf{x}} \times \nabla\left(\varepsilon_x^{-1}\hat{\mathbf{x}} \cdot \nabla \times \mathbf{H}\right) + \hat{\mathbf{y}} \times \nabla\left(\varepsilon_y^{-1}\hat{\mathbf{y}} \cdot \nabla \times \mathbf{H}\right) + \hat{\mathbf{z}} \times \nabla\left(\varepsilon_z^{-1}\hat{\mathbf{z}} \cdot \nabla \times \mathbf{H}\right) + k_0^2 \mathbf{H} = 0 \quad . \mathrm{A}(3)$$

For TM polarization, we take $\mathbf{H} = \hat{\mathbf{z}}h$, $\varepsilon_0 \mathbf{E} = \mathbf{d} = \hat{\mathbf{x}}d_x + \hat{\mathbf{y}}d_y$, here $d_y = -i\omega^{-1}\varepsilon_y^{-1}h^{(x)}$ and,

$$d_x = i\omega^{-1}\varepsilon_x^{-1}h^{(y)}. \quad \mathrm{A}(4)$$

Then, employing the normalized coordinates ($k_0 x \rightarrow x$, $k_0 y \rightarrow y$, $k_0 z \rightarrow z$, and $k_0 d \rightarrow d$), we arrive at the scalar wave equation:

$$h^{(x,x)} + \varepsilon_y\left(\varepsilon_x^{-1}h^{(y)}\right)^{(y)} + \varepsilon_y h = 0, \quad \mathrm{A}(5)$$

where $f^{(q)} = \partial f / \partial q$.

After using $\gamma_i = \varepsilon_{xi}^{-1} g_i = \left[\varepsilon_{yi}^{-1}\varepsilon_{xi}^{-1}k_x^2 - \varepsilon_{xi}^{-1}\right]^{1/2}$, we have,

$$\begin{aligned} &e^{2\gamma_2 \varepsilon_{x2} d}\left(\gamma_1 + \gamma_2\right)\left(\gamma_2 + \gamma_3\right) - \\ &e^{-2\gamma_2 \varepsilon_{x2} d}\left(\gamma_1 - \gamma_2\right)\left(\gamma_2 - \gamma_3\right) = 0 \end{aligned}, \mathrm{A}(6)$$

or in another form the above reads as,

$$\tanh\left(2\varepsilon_{x2}\gamma_2 d\right) = -\gamma_2 \frac{\gamma_1 + \gamma_3}{\gamma_2^2 + \gamma_1 \gamma_3}. \quad \mathrm{A}(7)$$

This equation can be simplified when the upper cladding $A_1$ and the lower cladding $A_3$ are made of the same anisotropic material ($\gamma_1 = \gamma_3$):

$$\tanh\left(2\varepsilon_{x2}\gamma_2 d\right) = -\frac{2}{\gamma_2/\gamma_1 + \gamma_1/\gamma_2}. \quad \mathrm{A}(8)$$

From trigonometric identities we obtain,

$$\tanh(\gamma_2 \varepsilon_{x2} d) = -\gamma_1/\gamma_2, \quad \mathrm{A}(9)$$

or,

$$\tanh(\gamma_2 \varepsilon_{x2} d) = -\gamma_2/\gamma_1. \quad \mathrm{A}(10)$$

When a symmetric 1-D waveguide is made of isotropic materials, Eqs. A(9) and A(10) degenerate into,

$$\tanh(g_2 d) = -g_1 \varepsilon_2 \left(g_2 \varepsilon_1\right)^{-1}, \quad \mathrm{A}(11)$$

$$\tanh(g_2 d) = -g_2 \varepsilon_1 \left(g_1 \varepsilon_2\right)^{-1}. \quad \mathrm{A}(12)$$

where $\varepsilon_1$ and $\varepsilon_2$ denote the dielectric function of the cladding and the core, respectively. Equations A(11) and A(12) match known results, see for example [29].

For TE polarization, there is no solution for plasmonic modes as long as all the materials are non-magnetic, which is the case for natural materials in optical range.

Equations A(11) and A(12) can be easily modified for photonic modes. For TM polarization, $g_2$ in Eqs. A(11) and A(12) should be replaced with $ig_2$ and $ig_1$, respectively, for waves guided in the dielectric core, in contrast to the case of plasmonic modes (SPPs) guided along the core-claddings interfaces. Using the relation, $i\tanh ix = -\tan x$, we arrive at the following DEs, defining the TM-polarized photonic modes

$$\tan(g_2 d) = -g_2 \varepsilon_{x1} \left( g_1 \varepsilon_{x2} \right)^{-1}, \quad \text{A(13)}$$

$$\tan(g_2 d) = g_1 \varepsilon_{x2} \left( g_2 \varepsilon_{x1} \right)^{-1}, \quad \text{A(14)}$$

where $g_2 = \left[ \varepsilon_{x2} - \varepsilon_{x2} \left( \varepsilon_{y2} \right)^{-1} k_x^2 \right]^{1/2}$.

For TE polarization, we take advantage of the duality relations between electric and magnetic parameters. The dielectric constants in Eqs. A(13) and A(14) ($\varepsilon_{x1}$, $\varepsilon_{y1}$, $\varepsilon_{x2}$, and $\varepsilon_{y2}$) are replaced by their counterparts, $\mu_{x1}$, $\mu_{y1}$, $\mu_{x2}$, and $\mu_{y2}$, which are all equal to one in optical range. Hence, the DEs for the TE-polarized photonic modes are,

$$\tan(k_2 d) = -k_2 / k_1, \quad \text{A(15)}$$

$$\tan(k_2 d) = k_1 / k_2, \quad \text{A(16)}$$

where, $k_2 = \left[ \varepsilon_{x2} - k_x^2 \right]^{1/2}$.